\documentclass[11pt]{iopart}
\usepackage{iopams}
\usepackage{graphicx}
\usepackage{mathrsfs}
\begin{document}
\bibliographystyle{unsrt}
\maketitle
\title[Scaling of Ergodicity in Binary Systems]{Scaling of Ergodicity in Binary Systems}

\author{M. S\"uzen}
\address{AMInstP \\ A89, Assia (Pa{\c s}ak\"oy) 5561 Famagusta, Cyprus}
\ead{mehmet.suzen@physics.org}

\begin{abstract}
Given pseudo-random binary sequence of length $L$, assuming
it consists of $k$ sub-sequences of length $N$. We estimate
how $k$ scales with growing $N$ to obtain a {\it limiting} ergodic
behaviour, to fulfill the basic definition of ergodicity (due to
Boltzmann). The average of the consecutive sub-sequences plays the role
of time (temporal) average. This average then compared to ensemble average
to estimate quantitative value of a simple metric called Mean Ergodic Time (MET),
when system is ergodic.
\end{abstract}

\section{Introduction}

Time averages play a central role in physics and statistical
mechanics. An ergodic theorem provides a programme to compute
ensemble averages of equilibrium properties of many
component dynamics equivalently by taking time averages in the
state-space \cite{tolman,farquhar}. Hence, it is important
to know when a dynamical system behave ergodically
i.e. how long does it take to approach to the ergodic
limit \cite{neirotti00a}, in order to reliably use ergodic
theorem and determine the averaged property in system where
finding ensemble averages is not computationally feasible or 
not known {\it a priori}. Even finding ensemble averages is possible,
covering the whole state-space may not be meaningful at all for the
observable in consideration.

A particular system, the binary system is used to model many different
physical or computational systems. While, it is quite simple in
construction, measuring the ergodicity in this system via generating 
a large pseudo-random binary sequences essentially provides quantitative
scaling measure.

\section{Ensemble Average of a Binary System}
\label{bs}

Consider a binary sequence ${a_{j}}, j=1,..,L$, each value in the
sequence representing a state or a point in the phase-space, that can 
take two values $a_{j} \in {0,1}$, similar to a spin system \cite{kittel}.

For example, a set of ensemble of two state ($N=2$) binary system simply takes the following form:  
$\mathscr{S}=(00,11,10,01)$, and a binary sequence representing this is simply $00111001$. The length 
of this sequence $L$ is $8$, and the number of sub-sequences of length $N=2$ is $k=4$
(size of the set $\mathscr{S}$).

In general, using the values in each state indexed $i=1,..,N$, over different ensemble
elements indexed $l=1,..,k$ (members within the set $\mathscr{S}$), {\it ensemble 
average} of each state $\langle A_{i} \rangle$ can be determined as follows:
\begin{equation}
 \label{astate}
  \langle A_{i} \rangle = \frac{1}{k} \sum_{l=1}^{k}  a_{l}^{i}.
\end{equation}
Recall that $L, N, j, i, k \in \mathbb{Z}^{+}$ and for any state this ensemble average must
be $1/2$ for a binary systems.

\section{Measure of the Mean Ergodicity Time}
\label{measure}

If we generate a very long pseudo-random sequence that represents a temporal evolution (data) of
the system, the average of each state is determined by a similar expression given in 
Equation (\ref{astate}), we call this value the {\it time average}. The estimate of the 
time (the number of sub-sequences, $k$) needed to reach vanishing difference in between 
the {\it time average} and {\it ensemble average} is called here {\it the mean ergodicity time} (MET)
as a simple metric.

Since we are doing numerical experiments, defining a target metric of this difference is
convenient. This metric $D_{MET}$ is defined as follows.
\begin{equation}
 \label{dergo}
  D_{MET} = \left| \langle A_{i} \rangle^{ensemble} - \langle A_{i} \rangle^{time} \right|.
\end{equation}
When the difference reach to a vanishing value we record the number of sub-sequence
($k$ above) visited. Repeating this procedure will generate an estimate for the 
MET.

Using a reliable pseudo-random sequence is critical in determining averages of given number of 
states (sub-sequence blocks) in the ensemble (the whole sequence). We have used 
Mersenne Twister (MT) \cite{matsumoto98a} that has a super astronomical period 
within the Maxima package \cite{maxima} to generate a binary sequence. The measure of randomness 
and its quality in generated pseudo-random sequences discussed elsewhere in detail \cite{compagner91,janke02b}.

\begin{table}[th]
  \begin{tabular}{|c|r|r|c|}
   \hline 
   {\bf $N$ state} &      {\bf   Mean Ergodicity Time (MET) }&   {\bf  Error}     & {\bf Number of Measurement} \\
      1       &                $685.5$          &   $343.41$   &  $60$  \\
      2       &                $29208.0$        &   $4416.15$  &  $373$ \\
      3       &                $194768.0$       &   $27747.53$ &  $87$  \\
      4       &                $249470.0$       &   $32428.54$ &  $56$  \\
      5       &                $343256.0$         &   $36243.63$ &  $70$  \\
      6       &                $517802.0$         &   $70089.40$ &  $33$  \\
   \hline 
  \end{tabular} \\
\caption{Scaling of MET with increasing $N$ state. Number of Measurements mean the number of different
        ergodicaly behaving segments within the generated random sequence. Average length of these
        segments are defined to be MET.}
  \label{data_met}
\end{table}

The above procedure is nothing but to fulfill the basic definition of ergodicity. 
The computation of {\it ensemble mean values} of thermodynamic observable ($U$) \cite{tolman} are 
multidimensional integrals over configuration space ${\bf x}$  \cite{neirotti00a},  
$$ \langle U \rangle = \int d {\bf x} P ( {\bf x} ) U( {\bf x} ) $$
where $P ( {\bf x} )$ is the probability of finding a system in the configuration
${\bf x}$.  In principle, the arithmetic mean value $\bar{U}$, that is averaged over 
temporal evolution, must be equal to its {\it ensemble mean value} for a system 
in ergodic behaviour. The analogy of this definition for a binary system is 
given above.

\begin{figure}[ptb]
\begin{center}
\includegraphics[width=0.5\columnwidth]{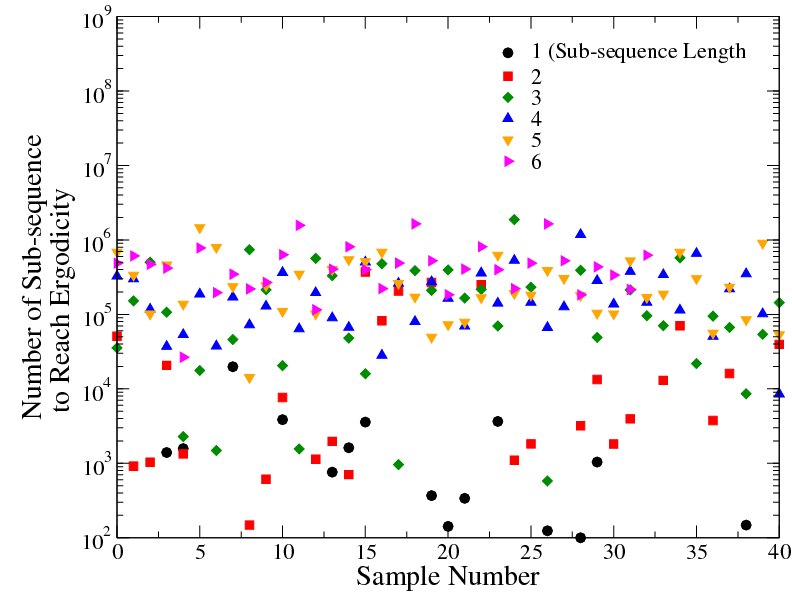}
\end{center}
\caption[Limiting Length]{The sketch of measurements of the number of consecutive sub-sequences to reach 
 ergodicity, the points where $D_{MET}$ metric reaches to value of $10^{-3}$.}
\label{samples16}
\end{figure}

\section{Scaling of Ergodicity}
\label{scale}

We have measured number of sub-sequences ($k$) with increasing number of states (up to $N=6$). 
The measured MET and other details of the data is given in Table (\ref{data_met}) and
in Figure (\ref{samples16}).

\begin{figure}[ptb]
\begin{center}
\includegraphics[width=0.5\columnwidth]{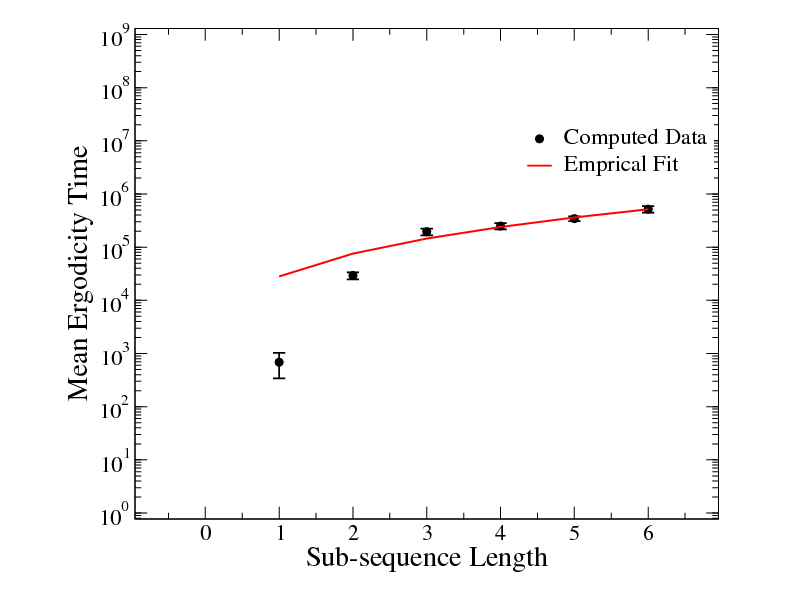}
\end{center}
\caption[Scaling]{Scaling of ergodicity against the number of binary states (mimics
the size of the state-space), averaged over many measurements summarized in Table (\ref{data_met}), 
fitted to an empirical function curve given in Equation (\ref{met}). }
\label{scale16}
\end{figure}

The scaling of MET against increasing number of states ($N$) is fitted in to an empirical 
law:
\begin{equation}
\label{met}
 MET = \alpha \exp(\beta N^{\gamma}).
\end{equation}
The scaling is shown in Figure (\ref{scale16}). Where the value of scaling coefficients 
are found to be $\alpha=87.16$, $\beta=5.77$ and $\beta=0.227$. This scaling can be 
used to check when a binary system with large number of states reach to an 
ergodic behaviour.

\section{Conclusions}
\label{concl}

A simple measure of scaling of ergodicity in binary systems, the length needed to 
reach ergodic behaviour with increasing system size, is explored by using 
a reliable pseudo-random binary sequence.  Our result would help to determine a 
lower bound on how long an experiment on a binary system should be repeated until 
it reaches to a point to make averaging that is thermodynamically acceptable. The 
work presented here can also be used as a pedagogical tool in 
understanding ergodicity of a dynamical system.

\appendix
 \section{Algorithm of MET measurements: Maxima Code}
Here we present the Maxima code to generate the results.
Note that given comments with a // are not valid Maxima syntax
and must be removed in the actual code.
 \begin{verbatim}
time_average(N,tolarence) := ( // N is the state size 
    sub_unit:[], // initialize system 
    sub_ave:[],
    for i:1 thru N step 1 do   
    (
      sub_unit:append(sub_unit,[0]),  // assign dummy values initially
      sub_ave:append(sub_ave,[0])
    ),
      up:2, // initial upper bound for mean ergodicity time
    for i:1 thru up step 1 do
    (
     for j:1 thru N step 1 do
       (
         // populate system and collect for averages
           num:random(2), sub_unit[j]:sub_unit[j]+num 
       ),
        sub_ave[1]:bfloat(sub_unit[1]/i), // initial time averages
        diff:abs(bfloat(0.5)-sub_ave[1]), //  and check
     for k:2 thru N step 1 do  // find out maximum difference 
                               // between time average and ensemble averages
       (
        sub_ave[k]:bfloat(sub_unit[k]/i) ,
        tsub: sub_ave[k],
        difft:abs(bfloat(0.5)-sub_ave[k]), 
        if difft > diff then diff:difft
       ),
      if diff >= tolarence then up:up+1 // increment for MET 
                                       // otherwise algoritm stops
    ),
      print("",up), // Report MET
      print("#time_ave=",tsub) // and time average (must be close to 0.5)
);
 \end{verbatim}
   
 \bibliography{research}
\end{document}